\documentclass[%
 reprint,
 amsmath,amssymb,
 aps,
]{revtex4-1}

\usepackage{graphicx}
\usepackage{dcolumn}
\usepackage{bm}
\usepackage{hyperref}
\usepackage{siunitx}

\begin{document}


\title{Thermal transport and frequency response of localized modes on low-stress nanomechanical silicon nitride drums featuring a phononic bandgap structure}
\author{Pedram Sadeghi}
\author{Manuel Tanzer}
\author{Niklas Luhmann}
\author{Markus Piller}
\author{Miao-Hsuan Chien}
\author{Silvan Schmid}
 \email{silvan.schmid@tuwien.ac.at}
\affiliation{%
 Institute of Sensor and Actuator Systems, TU Wien, Gusshausstrasse 27-29, 1040 Vienna, Austria.
}%

\date{\today}

\begin{abstract}

Development of broadband thermal sensors for the detection of, among others, radiation, single nanoparticles, or single molecules is of great interest. In recent years, photothermal spectroscopy based on the shift of the resonance frequency of stressed nanomechanical resonators has been successfully demonstrated. Here, we show the application of soft-clamped phononic crystal membranes made of silicon nitride as thermal sensors. It is experimentally demonstrated how a quasi-bandgap remains even at very low tensile stress, in agreement with finite element method simulations. An increase of the relative responsivity of the fundamental defect mode is found when compared to that of uniform square membranes of equal size, with enhancement factors as large as an order of magnitude. We then show phononic crystals engineered inside nanomechanical trampolines, which results in additional reduction of the tensile stress and increased thermal isolation, resulting in further enhancement of the responsivity. Finally, defect mode and bandgap tuning is shown by laser heating of the defect to the point where the fundamental defect mode completely leaves the bandgap.

\end{abstract}

\keywords{}

\maketitle

\section{Introduction}

Photothermal sensing based on nanomechanical silicon nitride (SiN) resonators has demonstrated exceptional sensitivity for the detection of radiation \cite{Piller2019} and single molecules \cite{Chien2018}. The direct measurement of photothermally absorbed power via the detuning of the resonance frequency enables thermomechanically-limited low background noise and high responsivity. The possibilities of qualitative and quantitative analysis on a great variety of analytes ranging from complex chemical compounds \cite{Yamada2013,Kurek2017} to nanoparticles \cite{Larsen2013,Schmid2014} also indicate the applicability of this technique to different sample requirements.

Surrounding SiN mechanical resonators with phononic crystals (PnCs) has been shown to reduce acoustic radiation losses into the supporting silicon (Si) frame \cite{Yu2014,Tsaturyan2014}. Patterning the PnC directly into the resonator leads to so-called "soft-clamping" in addition to reduced radiation losses \cite{Tsaturyan2017}, which results in enhanced damping dilution and has lead to quality factors ($Q$s) approaching 1 billion at room temperature \cite{Tsaturyan2017,Ghadimi2017,Ghadimi2018}. Central focus of these studies has been to increase the $Q$. Since damping dilution improves with tensile stress, only stoichiometric high-stress ($>$ 1 GPa) SiN has been used. In contrast, for applications such as thermal sensing, silicon-rich low-stress ($<$ 200 MPa) SiN is of interest due to the inverse relation between tensile stress and thermal responsivity \cite{Schmid2016,Chien2018}. To our knowledge, no investigation of low-stress PnC structures has been done prior.

Here, we propose using low-stress soft-clamped PnC resonators for thermal sensing. Two effects are expected to enhance the thermal response. First, the addition of holes to a resonator reduces its effective thermal conductivity, which results in increased thermal responsivity. Numerous investigations into the effect of PnCs on the thermal transport and sensitivity of perforated resonators has been done prior, in particular for the case of Si \cite{Maldovan2013,Zen2014,Zhang2019,Takahashi2020}. Second, defect modes engineered into PnCs display a larger overlap of the mechanical mode to the temperature field compared to a regular membrane mode, as illustrated in Fig.~\ref{fig1}(a). By comparing the temperature profile of a uniform membrane and the displacement fields of the fundamental uniform membrane mode and defect mode, respectively, obviously a much larger overlap with the temperature profile can be observed for the defect mode shape. As a result one can expect a localized higher average temperature in the PnC defect, which additionally leads to an increased thermal response. The chosen PnC design for all studied samples is taken from Tsaturyan et al., since this is the first and most studied soft-clamped PnC system \cite{Tsaturyan2017}.

Focus will be on various aspects of low-stress PnC samples. First, we show through numerical simulations how the bandgap varies with tensile stress and compare with experimental data. The focus then shifts to the thermal response of PnC membranes and how these perform relative to uniform membranes without any PnCs. We show for the first time a PnC engineered into a trampoline resonator, and how the thermal sensitivity is enhanced as a result. Finally, given the large overlap between the temperature and displacement fields of the defect, we perform defect mode and bandgap tuning of the PnC membranes through laser heating, demonstrating that it is possible to detune the defect mode completely out of the bandgap.

\section{Methods}

\begin{figure*}
  \centering
  \includegraphics[width=\textwidth]{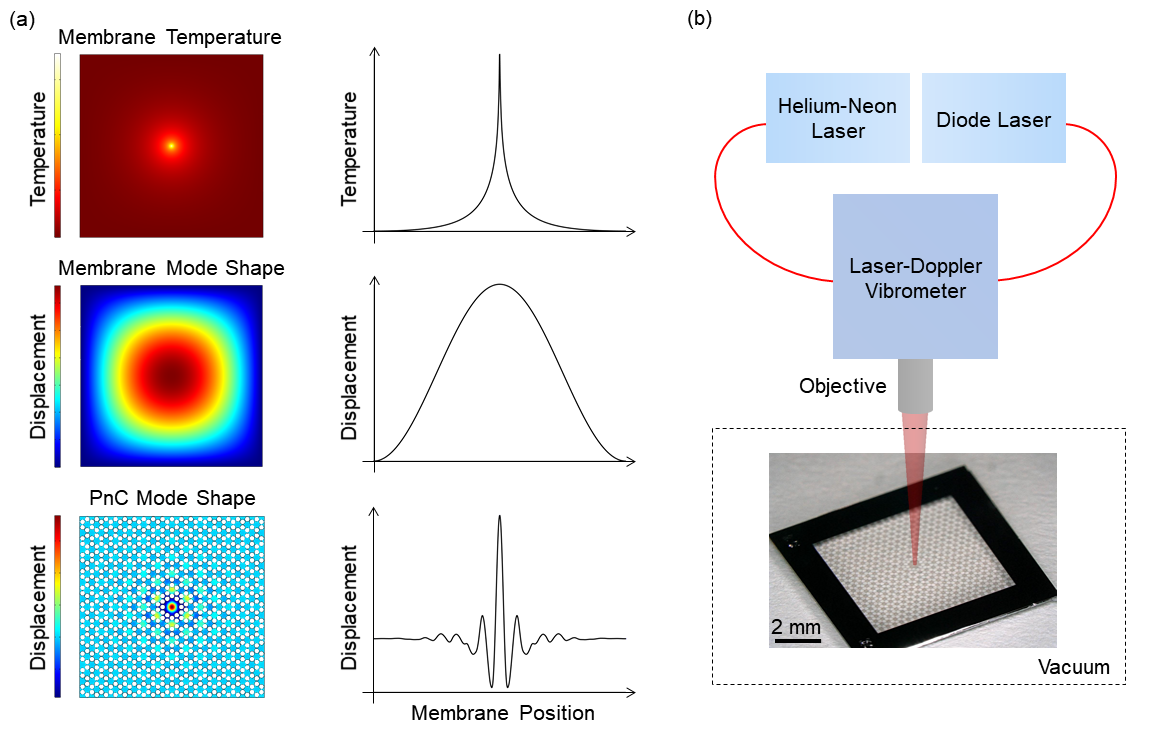}
  \caption[Figure 1]{Schematics of the overlap between temperature and displacement fields and the measurement setup. (a) Temperature profile of a uniform membrane compared to the displacement field of both a uniform membrane and a PnC membrane. The line plots are made along the diagonal of the membranes. Note that the line plot of the PnC displacement is a polynomial fit through the simulated values for clarity. (b) The thermal detuning of the resonance frequency is measured using a laser-Doppler vibrometer. A Helium-Neon laser is used to read out the vibration, while a secondary diode laser is used for the photothermal tuning of the frequency. A photograph of a single chip with a PnC membrane is shown as well. Samples are placed inside a high vacuum chamber in order to minimize gas damping and convective heat transfer.}\label{fig1}
\end{figure*}

All experiments are conducted with \SI{50}{\nano\meter} thin silicon-rich SiN deposited on \SI{380}{\micro\meter} Si wafers using low pressure chemical vapour deposition (LPCVD). A \SI{100}{\nano\meter} thick wet oxide (WOx) layer was grown on the Si wafer before SiN deposition, which acts as a sacrificial layer during backside release. The phononic crystal shapes are structured by reactive ion etching (RIE) with CF$_4$ down to the WOx layer. The backside openings for the final release with potassium hydroxide (KOH) are structured likewise but including the WOx layer, using combined RIE with CHF$_3$ and O$_2$. Using KOH, both the Si and WOx layers are etched from the backside and rinsed using deionized water.

Analysis of the mechanical properties is conducted with a commercial laser-Doppler vibrometer (MSA-500 from Polytec GmbH) equipped with a Helium-Neon laser with a center wavelength $\lambda = 633$ nm. Thermal responsivities are extracted by measuring the shift in the thermomechanical noise peaks for increasing laser powers. Therefore, a secondary diode laser (LPS-635-FC from Thorlabs GmbH), with a center wavelength $\lambda = 638$ nm, is used for varying power on top of the fixed vibrometer laser power. For both lasers, the exact power values are recorded using a silicon photodiode (S120C from Thorlabs GmbH). All vibrational experiments are performed under high vacuum (pressure p $\sim$ $10^{-6}$ mbar) in order to minimize gas damping and reduce heat transfer to the surrounding gas \cite{Verbridge2008}. A schematic of the experimental setup is given in Fig.~\ref{fig1}(b), including a photograph of a PnC membrane chip.

\begin{figure*}
  \centering
  \includegraphics[width=\textwidth]{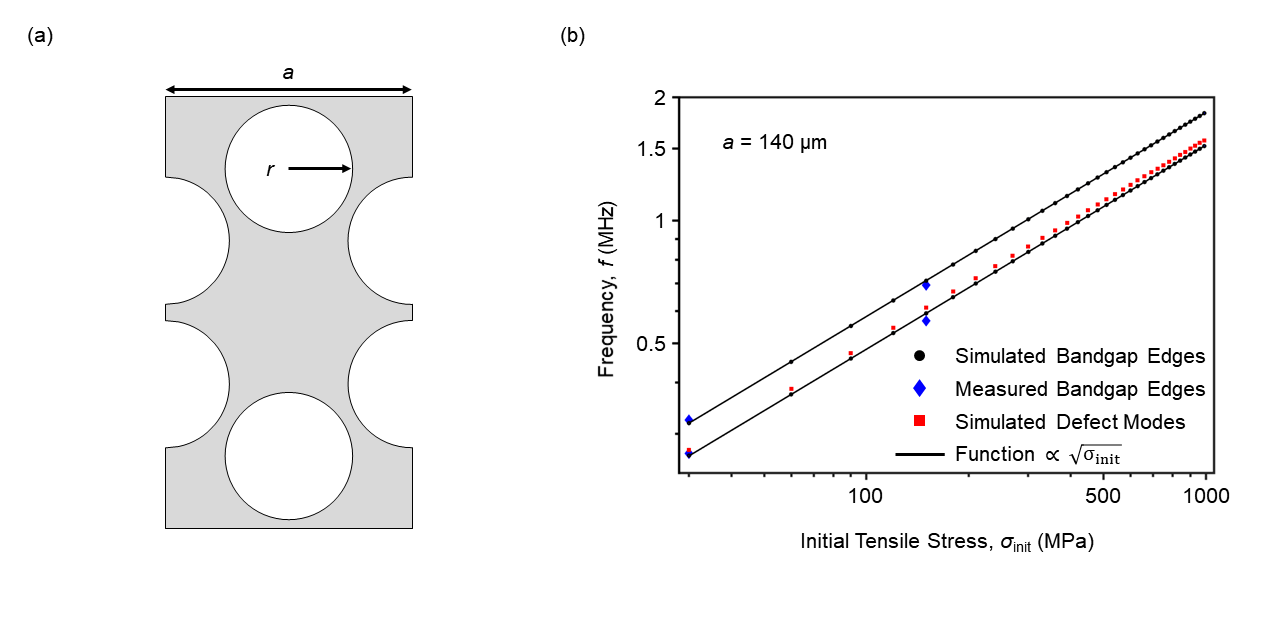}
  \caption[Figure 2]{Phononic bandgaps for low-stress SiN. (a) A single unit cell of the phononic crystal design used for all samples, similar to that used by Tsaturyan et al. \cite{Tsaturyan2017}. (b) Influence of initial tensile stress $\sigma_{\text{init}}$ on the width and center frequency of the phononic bandgap for a lattice constant $a = 140$ \si{\micro}m. Finite-element method simulations of the bandgap edges are given as black dots, while the full lines are fits to a function proportional to $\sqrt{\sigma_{\text{init}}}$. Measured bandgap edges are given as blue diamonds. Red squares represent FEM simulations of the fundamental defect mode frequency.}\label{fig2}
\end{figure*}

Simulations of the phononic bandgaps of PnC membranes are performed using COMSOL Multiphysics Version 5.5. Given the large aspect ratios of the samples, the Shell interface of the Structural Mechanics module is employed. Assumed material parameters are Young's modulus $E = 250$ GPa, mass density $\rho = 3000$ kg/m$^3$, and Poisson's ratio $\nu = 0.23$. In order to find the (quasi) phononic bandgap, a single unit cell is simulated. As a first step, the stress distribution in the unit cell is simulated assuming a biaxial initial tensile stress $\sigma_{\text{init}}$ taken from experimental measurements of the resonance frequency, while the boundaries are kept fixed (zero displacement). In the second step, using the simulated stress distribution from step 1, an eigenfrequency study is performed employing periodic boundary conditions instead of fixed boundaries in order to determine the phononic bandgap. Fig.~\ref{fig2}(a) shows a schematic of a single unit cell used for the bandgap simulations. The relation between lattice constant, $a$, and hole radius, $r$, is kept constant for all samples and given as $r = 0.26 a$, similar to Tsaturyan et al. \cite{Tsaturyan2017}. The lateral size $L$ of the square SiN membrane featuring the PnC was always scaled with the respective lattice constant $a$ with the fixed ratio $L \sim 19.4 a$.

Simulations of the defect modes are similarly done in a two step manner, but instead simulating the entire finite-sized PnC membrane. For this simulation, the boundaries are kept fixed for both the stationary and eigenfrequency simulations. Temperature distributions are simulated using the whole PnC membrane as well, but additionally employing the 'Heat Transfer in Shells' module of COMSOL. Laser heating is simulated using a Gaussian beam profile with a given power located on the central defect. Room-temperature boundary conditions are set at the membrane edges. The simulations are arranged such that heat can be transferred via conduction to the edge or via radiation to the environment. Thermal responsivities are extracted by simulating the eigenfrequency detuning, $\Delta f$, for five different power steps. For all FEM simulations, the maximum mesh size was made increasingly fine until the simulated bandgap width, center frequency, and defect mode frequency would change by less than 10$\%$ upon further mesh refinement.

\section{Results and Discussion}

\subsection{Effect of tensile stress on PnC bandgap}

Since the thermal responsivity is inversely proportional to the effective tensile stress, $\sigma_{\text{eff}}$, of the SiN resonator, it is central to the application of PnC membranes as thermal sensors to investigate bandgap width for low-stress resonators. FEM simulations of bandgap width versus the initial tensile stress $\sigma_{\text{init}}$ are given in Fig.~\ref{fig2}(b) for a PnC with a lattice constant $a = 140$ \si{\micro}m. For a uniform square membrane, one can assume $\sigma_{\text{eff}} = \sigma_{\text{init}}$, but given the holes of the PnC membranes, $\sigma_{\text{eff}}$ will be non-uniform and the two stresses have to be clearly distinguished. It can be seen that both the width of the bandgap and its center frequency scale with $\sqrt{\sigma_{\text{init}}}$, which is the expected behavior for the eigenfrequency of structures under tensile stress such as membranes or strings \cite{Schmid2016}. In addition to the bandgap, the resonance frequency, $f_r$, of the fundamental defect mode is shown as well, in order to highlight a similar relation to the initial stress and that it remains in the bandgap for all $\sigma_{\text{init}}$ values. Overlaid on the simulations are measured bandgaps from samples with initial stresses of 30 MPa and 150 MPa. Good agreement is observed both in terms of bandgap width and center frequency. These results show that PnC membranes maintain a bandgap even for very low initial stresses and should therefore be possible to employ in thermal sensing applications.

\subsection{Responsivity of defect modes}

The next step is to show the effect of PnCs on the responsivity of membrane resonators and how these compare to the previously studied uniform square membranes \cite{Chien2018}. The responsivity here refers to the relative frequency shift, $\delta f = \Delta f/f_r$, per absorbed power, $P_{\text{abs}}$. Initial investigations into the thermal response of square membranes only assumed heat transfer via conduction. With these assumptions, using a Taylor series approximation of the responsivity of a circular membrane, the relative responsivity of a square membrane has been shown to be well approximated by \cite{Kurek2017}:
\begin{equation}
    \delta R = \frac{\delta f}{P_{\text{abs}}} \approx -\frac{\alpha E}{8 \pi \kappa h \sigma_{\text{eff}}} \left(\frac{2 - \nu}{1 - \nu} - 0.642\right).\label{eq:1}
\end{equation}
Here, $\alpha = 2.2\cdot10^{-6}$ $K^{-1}$ is the thermal expansion coefficient of SiN, $\kappa$ is the thermal conductivity, and $h = 50$ nm the membrane thickness. From this equation, no dependence on lateral dimensions are expected. However, recent investigations into heat transfer in SiN have shown a non-negligible contribution from emission of thermal radiation \cite{Zhang2020,Piller2020}. This effect becomes more pronounced as the lateral size of the membrane increases and results in a reduction of responsivities compared to that predicted using equation~(\ref{eq:1}). Given the size of the membranes studied here, which are on the order of several millimetres, emission of thermal radiation has to be taken into consideration. As such, all simulations take emission into account with an emissivity $\epsilon = 0.05$, calculated using optical data of 50 nm SiN membranes \cite{Luhmann2020}. Further investigation into the effect of thermal radiation on the responsivity of membrane resonators can be found in Piller et al. \cite{Piller2020}.

\begin{figure*}
  \centering
  \includegraphics[width=\textwidth]{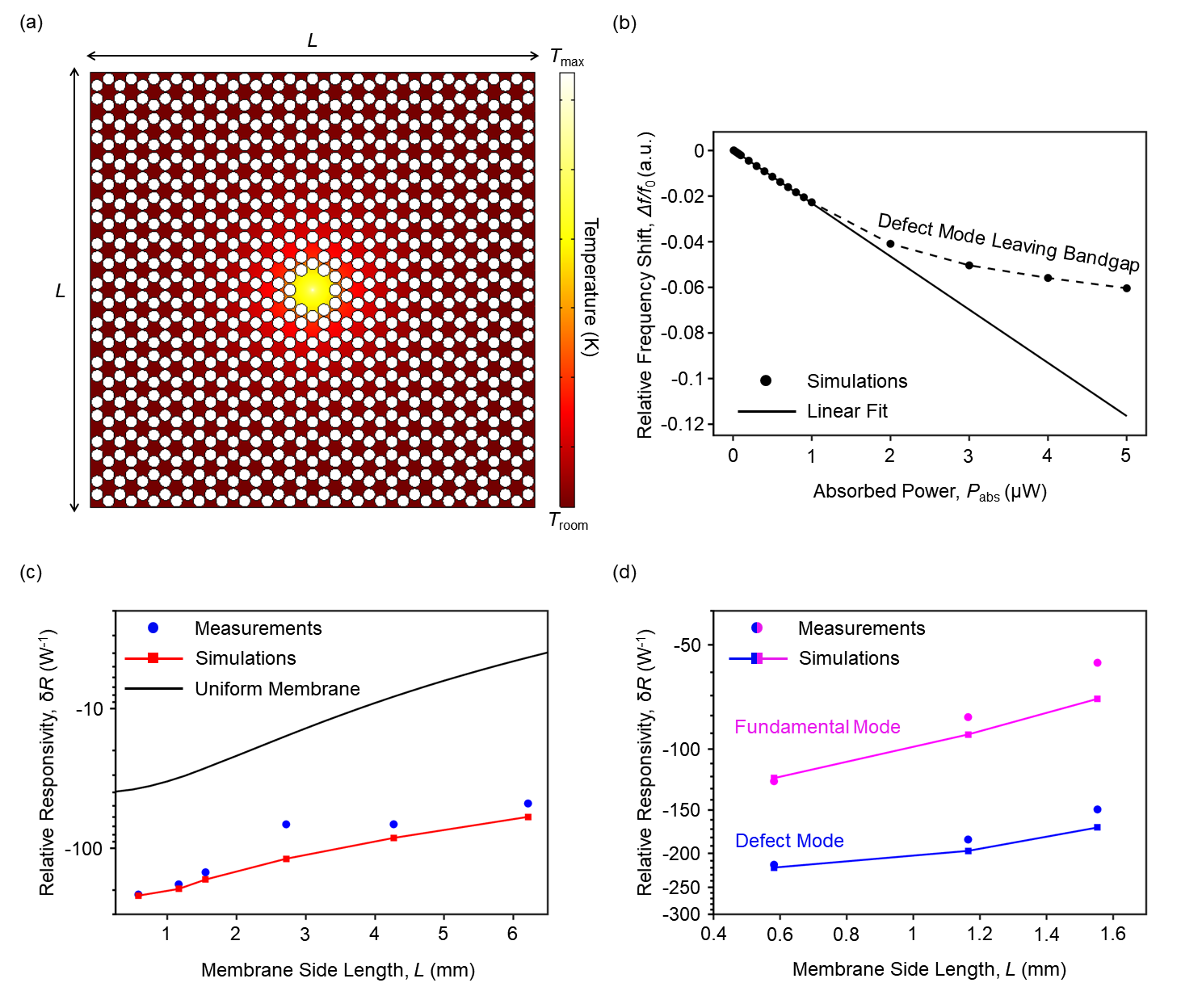}
  \caption[Figure 3]{Relative responsivities of 30~MPa phononic crystal membranes. (a) Temperature distribution plot of a PnC membrane with $L \sim 2.7$ mm and $P_{\text{abs}} = 0.6$ \si{\micro}W. The temperature at the center of the defect is $T_{\text{max}}$, while the boundary temperature is set to $T_{\text{room}} = 293.15$ K. (b) Simulated power dependence of the relative frequency shift for the membrane shown in (a). The dashed line is a guide to the eye. (c) Measured (blue dots) and simulated (red squares) relative responsivities of the fundamental defect mode of PnC membranes versus $L$. In addition, simulated responsivity values for uniform square membranes are given (black line). (d) Comparison between the fundamental defect modes (blue data) and the the fundamental out-of-plane mode of the entire PnC membrane (magenta data) versus $L$. Simulations for each set of modes are provided as well.}\label{fig3}
\end{figure*}

The motivation behind employing PnC membranes for thermal sensing can be illustrated through simulations of the temperature distribution when heat is absorbed in the defect center, as shown in Fig.~\ref{fig3}(a). Evidently, the temperature is highly localized to the central defect with a maximum temperature $T_{\text{max}}$, while quickly approaching room-temperature in the surrounding PnC.

Fig.~\ref{fig3}(b) shows simulations of the relative frequency shift of the fundamental defect mode for the membrane shown in Fig.~\ref{fig3}(a) as a function of the absorbed power, $P_{\text{abs}}$. A linear relation can be seen for powers $P_{\text{abs}} < 1$ \si{\micro}W, but start to deviate from linearity at larger powers. Similar simulations of uniform membranes of equal size were made, in which case the frequency shift remained linear for all power values shown in Fig.~\ref{fig3}(b) \cite{Piller2020}. As such, this nonlinearity turns out to be a result of a heating induced detuning of the fundamental defect mode to frequencies outside of the bandgap, which de-localizes the mode to the entire PnC membrane. Simulations of the defect mode shapes for higher values of $P_{\text{abs}}$ are shown in Supplementary Fig.~\ref{figS1}. For the following discussion, all responsivity values are extracted from the linear regime of the frequency shift.

Measurements of the relative responsivity for the fundamental defect mode of PnC membranes of various sizes are presented in Fig.~\ref{fig3}(c). FEM simulations of $\delta R$ are shown shown as well. In addition, simulated $\delta R$-values of the fundamental mode of uniform square membranes are shown. An enhancement of $\delta R$ for PnC membranes can be observed compared to the uniform membranes, with all $\delta R$-values being at least a factor of 5 greater than uniform membranes of equal size. As discussed above, $\delta R$ in the case of pure conduction-based thermal transport as modelled in equation~(\ref{eq:1}) would show no size dependence. The observed decrease in $\delta R$ with increasing $L$ is a result of the aforementioned radiative heat transport, which lowers the response with increasing membrane size. Importantly, the effect appears less pronounced for the PnCs than for uniform membranes, which is possibly a result of the smaller effective size of the defect mode compared to the fundamental mode of the entire membrane. Simulated $\delta R$-values for the PnCs show good agreement with measurements, indicating that thermal emission indeed is responsible for the size-dependence of $\delta R$.

In order to confirm that the observed $\delta R$-enhancement is a result of the better overlap of defect mode with the temperature field and not simply a result of a reduced $\kappa$ due to the holes engineered into the membrane, $\delta R$-values of the fundamental mode of the PnC membrane were measured as well. Fig.~\ref{fig3}(d) shows measured and simulated values of the $\delta R$ of the fundamental defect mode and the fundamental membrane mode. For all shown membrane sizes, the defect mode $\delta R$ is greater than that of the entire PnC membrane. Additionally, $\delta R$ appears to roll-off faster for larger membranes, possibly a result of the reduced overlap with the temperature field. Overall, the presented data show that both effects, namely reduced effective thermal conductivity and overlap of the temperature field to the displacement field, lead to a significant enhancement of the responsivity of defect modes when compared to square membranes of similar size. Additional data for higher order defect modes can be found in Supplementary Fig.~\ref{figS2}.

\subsection{Enhanced responsivity by geometrical stress reduction}
\begin{figure*}
  \centering
  \includegraphics[width=\textwidth]{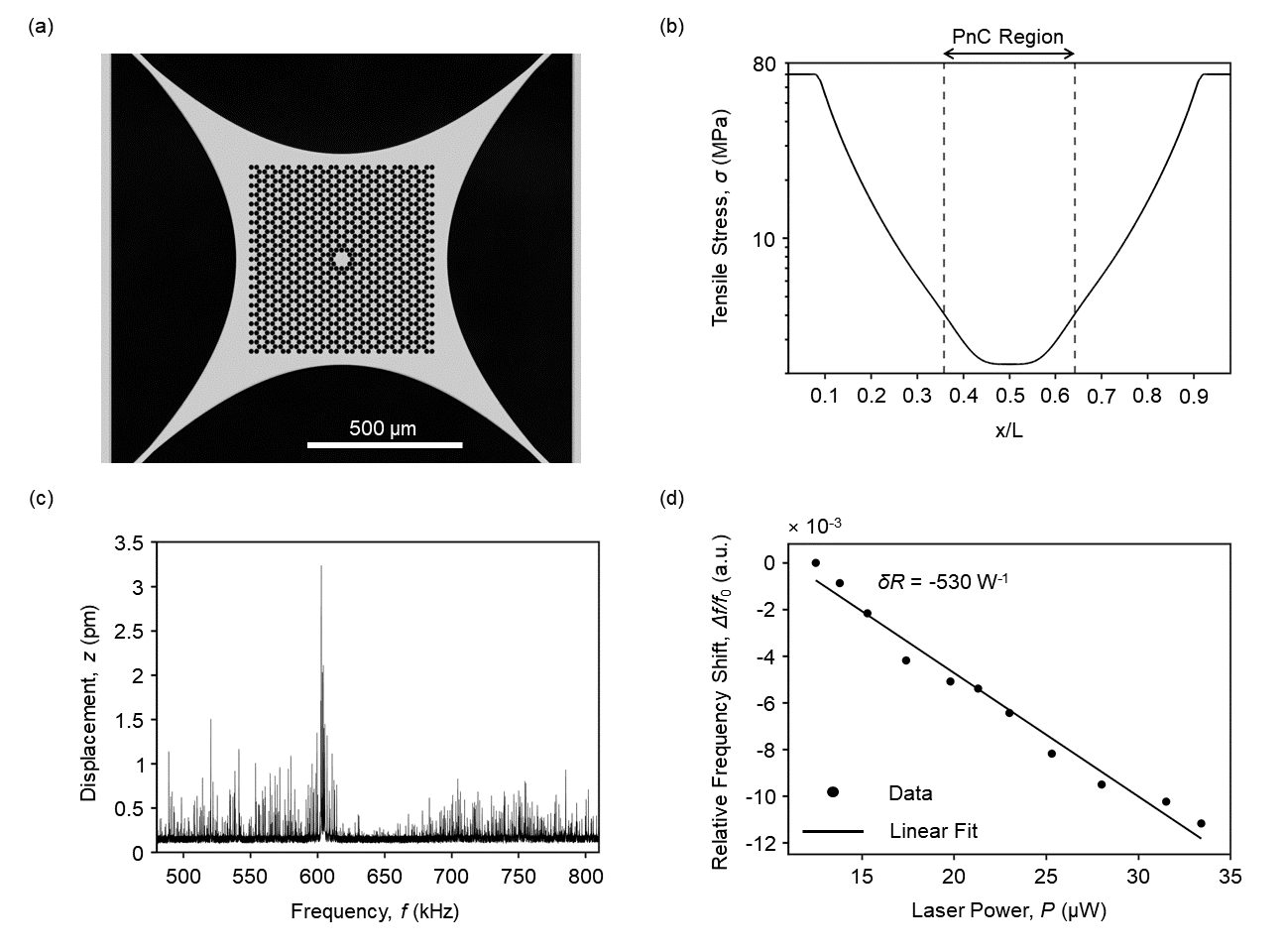}
  \caption[Figure 4]{Phononic crystal engineered into a nanomechanical trampoline. (a) Optical micrograph of a trampoline PnC with a lattice constant $a = 30$ \si{\micro}m and initial stress $\sigma_{\text{init}} = 30$ MPa. (b) Simulated tensile stress along the diagonal of the trampoline. (c) Measured Brownian motion of the fundamental PnC defect mode. (d) Relative frequency shift versus laser power for the fundamental defect mode of the trampoline PnC. A linear fit to the data is shown as well and a $\delta R = -530$ W$^{-1}$ can be extracted from the fit.}\label{fig4}
\end{figure*}

As discussed above in (\ref{eq:1}), reduced stress corresponds to increased responsivity. The next step would then be to further reduce the stress of PnC membranes in order to enhance $\delta R$. Given the difficulty in precisely controlling the tensile stress of SiN during the LPCVD process, alternative means have to be employed to reduce the stress. Previous studies have shown how this can be achieved using oxygen plasma \cite{Luhmann2017,Chien2018}. However, given how fragile PnC membranes are and the lack of precise control of the final tensile stress, the oxygen plasma method is not employed here. Attention is instead turned towards geometric strain engineering by designing PnCs inside trampoline resonators.

Nanomechanical trampolines have been studied previously in the field of quantum optomechanics \cite{Norte2016,Reinhardt2016}. The four narrow tethers of a trampoline ensures high stress localization to the tethers, while the central pad displays a much lower stress. Besides the stress engineering, the narrow tethers additionally improve the thermal isolation. These key features has led to trampoline resonators being used for thermal sensing applications, as demonstrated using silicon \cite{Varpula2017} and graphene \cite{Blaikie2019} as the trampoline material. Trampolines made of SiN have additionally been shown to work as position-sensitive detectors \cite{Chien2020}.

Here, we combine the reduced stress and increased thermal isolation of trampoline resonators with the demonstrated enhanced thermal response of PnCs by engineering the PnC directly into a trampoline. An optical micrograph of such a structure is given in Fig.~\ref{fig4}(a). The trampoline has an opening window size of 1.5~mm and a tether width of 20~\si{\micro}m at the corners of the window. As for the PnC, a lattice constant of $a = 30$ \si{\micro}m was chosen for this trampoline size. Fig.~\ref{fig4}(b) shows a simulation of the tensile stress along a diagonal line through a uniform trampoline without any PnC in the middle. It can be observed that the stress is significantly lower in the central pad compared to the tethers, as expected, which should translate into an increased responsivity of the PnC defect mode. The PnC region of the stress is highlighted on the plot.

A measurement of the Brownian motion of the trampoline PnC defect is given in Fig.~\ref{fig4}(c). A narrow and badly developed bandgap can be observed in the frequency range of 600-700~kHz. The fundamental PnC defect mode is observable at $f \sim 603$ kHz. However, unlike for PnCs engineered in square membranes, many peaks can be observed in the bandgap region, making it difficult to distinguish the exact boundaries of said bandgap. The lack of a well-defined bandgap can be attributed to the stress distribution in a trampoline as was shown in Fig.~\ref{fig4}(b). While the stress is much lower, it is not perfectly uniform throughout the central pad and thus the stress in the PnC varies with position.

\begin{figure*}
  \centering
  \includegraphics[width=\textwidth]{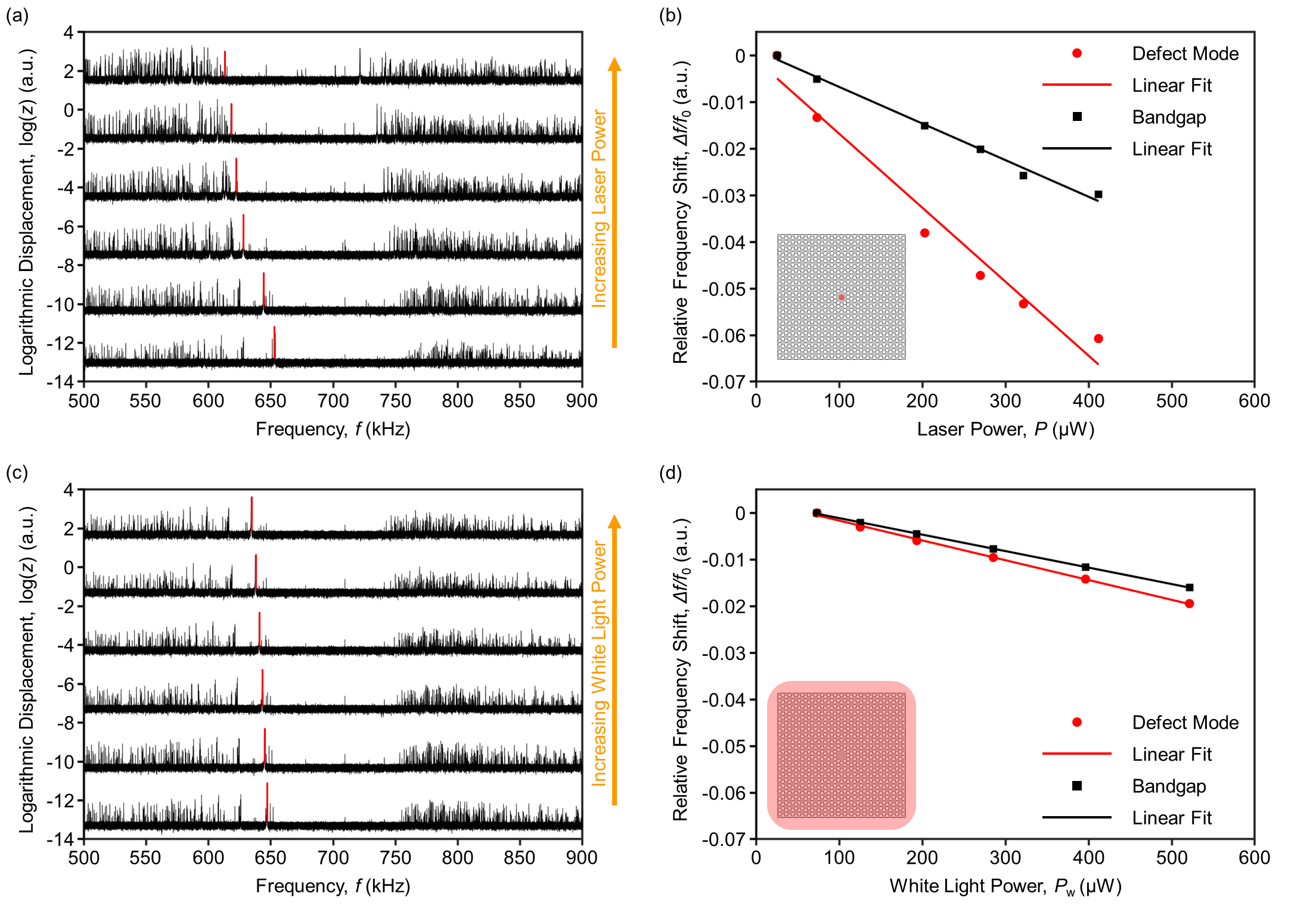}
  \caption[Figure 5]{Bandgap and defect mode tuning of phononic crystal membranes. (a) Logarithmic displacement versus frequency for a PnC membrane with $L \sim 1.2$ mm and $\sigma_{\text{init}} = 30$ MPa for increasing laser power. Displacement curves have been shifted vertically for clarity and the fundamental defect mode is highlighted in red. (b) Relative frequency shift of the fundamental defect mode (red dots) and the bandgap center frequency (black squares) for increasing laser power along with linear fits to each set of data. (c) Logarithmic displacement curves for increasing white light power. (d) Relative frequency shifts of the fundamental defect mode and bandgap center frequency for increasing white light power. Insets in (b) and (d) are schematics of laser and white light heating of the PnC, respectively.}\label{fig5}
\end{figure*}

Fig.~\ref{fig4}(d) shows the relative frequency shift for the fundamental defect mode of the trampoline resonator for increasing laser power focused on the defect center. From a linear fit to the data, a $\delta R = -530$ W$^{-1}$ can be extracted. Compared to a regular PnC membrane of equal lattice constant, an enhancement of the responsivity of $\sim$2.5 times is observed for the trampoline PnC, which represents the largest $\delta R$ observed among all PnC samples. The enhancement can be qualitatively understood from the reduced stress of the trampoline central pad: Using the stress plot in Fig.~\ref{fig4}(b), an average stress of $\sigma = 2.7$~MPa can be calculated in the region of the pad covering the PnC, significantly lower than the 30~MPa initial stress of the square PnC membranes. The lack of a linear enhancement, as expected from equation \ref{eq:1}, is most likely due to the uneven stress in the PnC. With precise and uniform stress reduction, even larger $\delta R$s are to be expected. Additionally, the tethers of the trampoline can be made narrower, which increases thermal isolation.

\subsection{Thermal defect mode and bandgap tuning}

The final part of this paper focuses on defect mode and bandgap tuning of the PnC membranes. As already shown from simulations in Fig.~\ref{fig3}(a), in the case of a central point source, the temperature is strongly localized to the central defect. Hence, the local thermal expansion of the defect mode causes a local reduction in stress that causes the defect mode frequency to drop. For large enough temperatures, the defect mode frequency can be detuned enough to exit the bandgap and start to de-localize into the entire PnC membrane, as was touched upon in Fig.~\ref{fig3}(b). Such defect mode tuning is shown in Fig.~\ref{fig5} for a regular PnC membrane without trampolines.

Fig.~\ref{fig5}(a) shows plots of the Brownian motion spectrum for increasing laser power focused on the center defect. Curves recorded at higher laser powers are shifted vertically for clarity. It can be observed how the defect mode (red) shifts faster as a response to the increasing laser power compared to the bandgap center frequency. This indicates a much larger temperature in the defect compared to the PnC. For the largest powers, the defect mode frequency approaches the lower edge of the bandgap and the peak amplitude drops to the level of the surrounding forest of modes, suggesting that the resonance has been shifted outside the phononic bandgap. This is better observed by plotting the relative frequency shifts of the defect mode and bandgap center frequency versus laser power, as can be seen in Fig.~\ref{fig5}(b). A linear fit describes the bandgap data well, but deviates slightly at larger powers for the defect mode data, similar to simulations shown in Fig.~\ref{fig3}(b).

In contrast to a heating point source on the center defect, the alternative is to uniformly heat the entire PnC structure, as shown Fig.~\ref{fig5}(c). The measurements are similar to those shown in Fig.~\ref{fig5}(a), but the PnC is illuminated with increasing power of the white light source of the vibrometer instead of the laser. For the chosen microscope objective, the white light covers the entire membrane, thus heating the whole structure equally. It can be observed that the defect mode and bandgap both shift slightly downward in frequency, but the location of the defect mode relative to the bandgap edge remains nearly constant. In this case, the temperature is no longer localized to the defect and both the defect mode and bandgap shift nearly equally, as discussed in Fig.~\ref{fig2}(b).

Fig.~\ref{fig5}(d) further highlights this fact by showing the relative frequency shifts for increasing white light power. Here, both the defect mode and the bandgap shift equally to a good approximation. The small difference can be attributed to the fact that the frequencies are still recorded using the vibrometer, thus having a constant laser heating on the central defect in addition to the white light heating.

\section{Conclusion}

In summary, we have demonstrated the application of soft-clamped phononic crystal membranes for thermal sensing. Measurements of the Brownian motion spectra show that a quasi-bandgap remains even for a very low initial tensile stress, in agreement with finite element method simulations. From FEM simulations, it can be shown that when the center defect is heated, the local temperature in the defect is much larger than the surrounding PnC. This temperature localization results in a larger overlap of the displacement field to the temperature field and thus in a potential enhancement of the relative responsivity by more than an order of magnitude. Through an experimental investigation of PnC membranes with various defect sizes, it is found that smaller defects generally result in larger responsivities because of the reduced cooling due to radiative thermal transport. Additional enhancements can be achieved by further reducing the tensile stress, which was shown here by embedding a PnC membrane into a nanomechanical trampoline. With this geometry, a relative responsivity $\delta R = -530$ W$^{-1}$ was achieved, which is a factor of 2.5 larger than the best value achieved with PnCs in square membranes.

We also demonstrate defect mode and bandgap frequency tuning of PnC membranes through laser heating of the central defect. It was found that the defect mode shows a stronger response to laser heating compared to the surrounding PnC, to the point that the mode was shifted outside of the bandgap. This was attributed to the much larger temperature in the defect compared to the PnC due to the temperature and displacement field overlap. Further confirmation of temperature localization was achieved through heating of the entire PnC membrane equally by use of the white light source of the vibrometer. In this case, the defect mode and PnC responded almost equally to increased light power. FEM simulations corroborated the experimental data very well.

Further control of the tensile stress reduction would potentially allow even higher $\delta R$s to be achieved than presented here. Addition of a suitable absorbing layer on top of the PnC presents an additional path towards greater responsivities \cite{Luhmann2020}. Even larger bandgaps can be opened through variation of the PnC design, which might lead to modes with even higher localization \cite{Reetz2019}. Overall, the data presented here show the promise of PnC membranes for thermal sensing applications and could pave the way for a new generation of broad range thermal sensors.


\bibliography{Lib}

\begin{acknowledgments}

The authors would like to acknowledge support from Sophia Ewert, Patrick Meyer, and Michael Buchholz with the sample fabrication. This work is supported by the European Research Council under the European Unions Horizon 2020 research and innovation program (Grant Agreement-716087-PLASMECS) and (Grant Agreement-875518-NIRD).

\end{acknowledgments}

\appendix
\renewcommand{\thefigure}{S\arabic{figure}}
\setcounter{figure}{0}

\section{Simulated Mode Shapes for Larger Laser Powers}

\begin{figure*}[ht!]
  \centering
  \includegraphics[width=\textwidth]{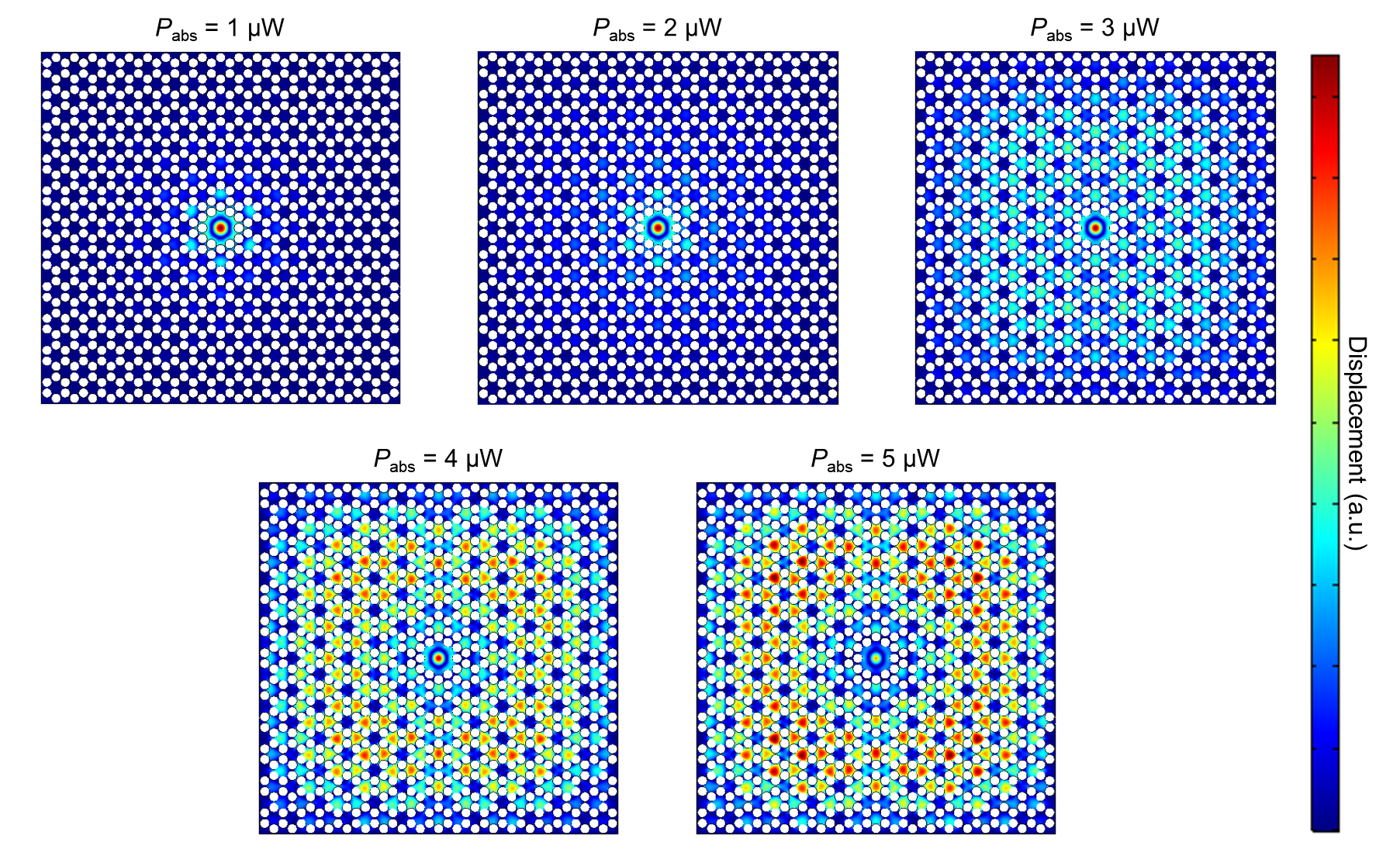}
  \caption[Figure S1]{FEM simulations of the fundamental defect mode shape for increasing absorbed power values.}\label{figS1}
\end{figure*}

Fig.~\ref{figS1} shows simulated mode shapes of the fundamental defect mode for the last 5 power steps in Fig.~\ref{fig3}(b). As a result of the mode slowly exiting the bandgap, the displacement field starts to penetrate further into the phononic crystal for each power step, until extending throughout the membrane for the largest power step.

\section{Responsivity of Higher Order Defect Modes}

\begin{figure*}[ht!]
  \centering
  \includegraphics[width=\textwidth]{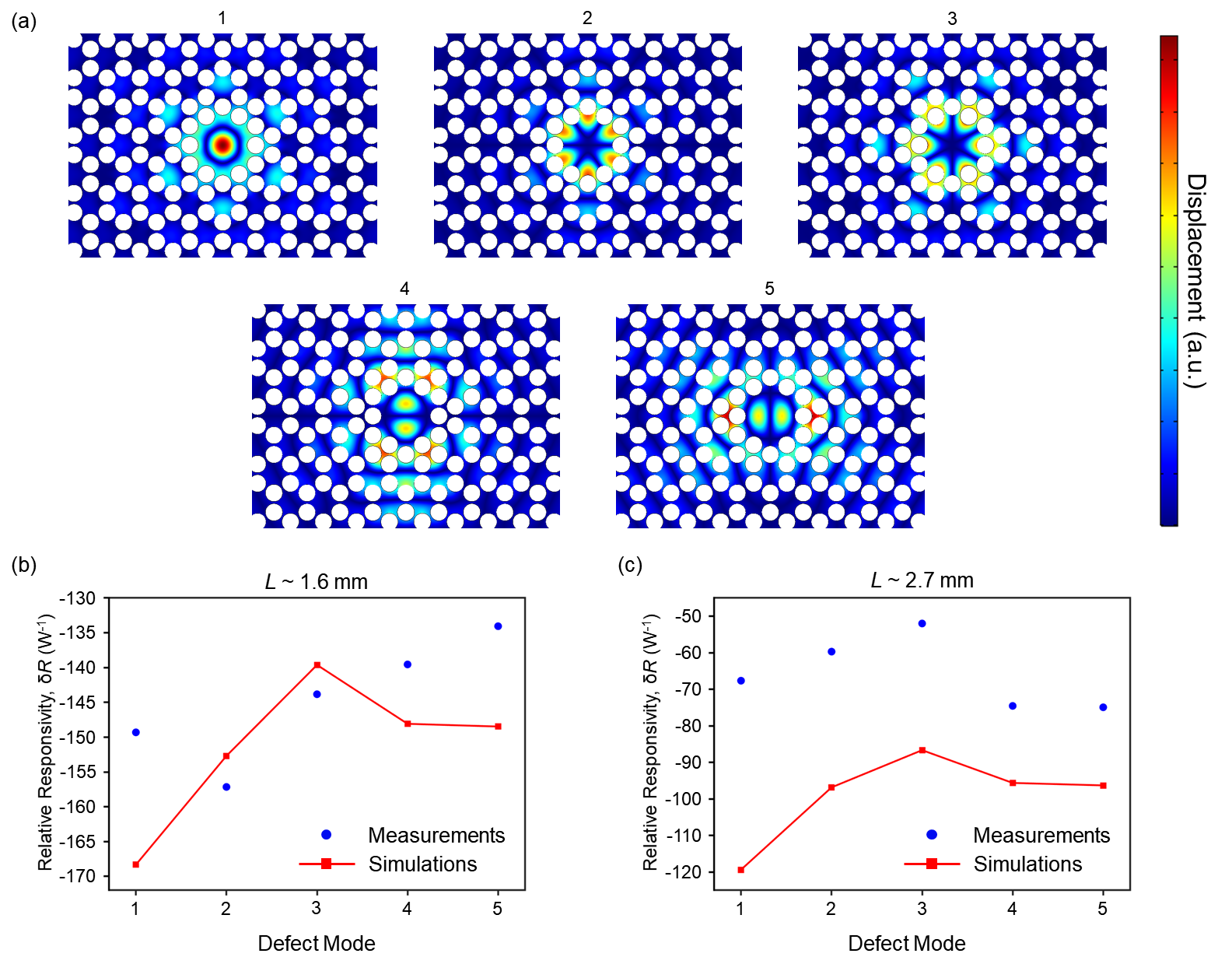}
  \caption[Figure S12]{Responsivities of higher order defect modes. (a) Simulated mode shapes of the five different defect modes observed and given the labels 1-5. (b) Measured (blue dots) and simulated (red squares) responsivites for the five defect modes of a phononic crystal membrane with a side length $L \sim 1.6$ mm. (c) Same as (b) for a membrane with $L \sim 2.7$ mm.}\label{figS2}
\end{figure*}

FEM simulations of the five different defect modes observed for this phononic crystal design are shown in Fig.~\ref{figS2}(a) and labeled as modes 1-5. Measured relative responsivities for the different defect modes are given in Fig.~\ref{figS2}(b) and (c) for membranes of size $L \sim 1.6$ mm and $L \sim 2.7$ mm, respectively. Values extracted from FEM simulations are shown as well. A clear trend cannot be observed when comparing the two sizes experimentally, possibly a result of the laser position drifting slightly over time during the experiments. However, from simulations it would appear responsivities follow a relation $1 > 4 \sim 5 \sim 2 > 3$. This pattern can possibly be explained from the power source in both experiments and simulations being placed at the center of the defect, which coincides with the antinode of the fundamental defect mode, but not with that of the higher order modes. If the power source was placed at the antinode of each mode, a more constant $\delta R$-value is to be expected.

\end{document}